\documentclass[sigconf, screen]{acmart}
\usepackage{graphicx,xspace,url} 
\usepackage{longtable,supertabular}
\usepackage{xcolor,textcomp}
\usepackage{hyperref}
\usepackage{booktabs}
\usepackage{pgfplots}          
\usepackage{breakurl}
\usepackage{subcaption}
\usepackage{pgfplots}
\usepackage{pgfplotstable}
\pgfplotsset{compat=1.9}
\usepackage{algorithm}
\usepackage{algpseudocode}
\usepackage{multirow}
\usepackage{verbatim}
\usepackage{enumitem}

\usepackage{letltxmacro}
\LetLtxMacro\oldttfamily\ttfamily
\DeclareRobustCommand{\ttfamily}{\oldttfamily\csname ttsize\endcsname}
\newcommand{\setttsize}[1]{\def\ttsize{#1}}%

\definecolor{ForestGreen}{rgb}{0.13, 0.55, 0.13}

\newcommand{\BW}{{\sc Birdwatch}\xspace}
\newcommand{\CR}{{\sc ClaimReview}\xspace}

\newcommand{\rev}[1]{\textcolor{black}{{#1}}}

\newcommand{\cam}[1]{\textcolor{black}{{#1}}}

\setcopyright{acmcopyright}
\copyrightyear{2022}
\acmYear{2022}
\acmDOI{10.1145/xxx}

\acmConference[CIKM'22]{CIKM'22: The ACM CIKM Conference}{2022}{Atlanta, GA, USA}
\acmBooktitle{CIKM'22: The ACM CIKM Conference, 2022, Atlanta, GA, USA}
\acmPrice{15.00}
\acmISBN{978-1-4503-XXXX-X/XX/XX}

\author{Mohammed Saeed}
\affiliation{
\institution{Eurecom}
\city{Biot}
\country{France}
 }
\email{mohammed.saeed@eurecom.fr}

\author{Nicolas Traub}
\affiliation{
\institution{The University of Queensland}
\city{Brisbane}
\country{Australia}
 }
\email{n.traubdamico@uq.net.au}

\author{Maelle Nicolas}
\affiliation{
\institution{Eurecom}
\city{Biot}
\country{France}
 }
\email{Maelle.Nicolas@eurecom.fr}

\author{Gianluca Demartini}
\affiliation{
\institution{The University of Queensland}
\city{Brisbane}
\country{Australia}
 }
\email{demartini@acm.org}

\author{Paolo Papotti}
\affiliation{
\institution{Eurecom}
\city{Biot}
\country{France}
 }
\email{paolo.papotti@eurecom.fr}

\begin{document}
\title{Crowdsourced Fact-Checking at Twitter: \\How Does the Crowd Compare With Experts?} 

\renewcommand{\shortauthors}{Saeed et al.}

\begin{abstract}
Fact-checking is one of the effective solutions in fighting online misinformation. However, traditional fact-checking is a process requiring scarce expert human resources, and thus does not scale well on social media because of the continuous flow of new content to be checked. Methods based on crowdsourcing have been proposed to tackle this challenge, as they can scale with a smaller cost, but, while they have shown to be feasible, have always been studied in controlled environments. In this work, we study the first large-scale effort of crowdsourced fact-checking deployed in practice, started by Twitter with the Birdwatch program. Our analysis shows that crowdsourcing may be an effective fact-checking strategy in some settings, even comparable to results obtained by human experts, but does not lead to consistent, actionable results in others. \rev{We processed 11.9k tweets verified by the Birdwatch program} and report empirical evidence of i) differences in how the crowd and experts select content to be fact-checked, ii) how the crowd and the experts retrieve different resources to fact-check, and iii) the edge the crowd shows in fact-checking scalability and efficiency as compared to expert checkers.

\end{abstract}

\maketitle

\section{Introduction}
\label{sec:introduction}
The spread of online misinformation carries risks for the democratic process and for a decrease in public trust towards authoritative sources of news \cite{starbird2019disinformation}. 
%
Fact-checking is one of the prominent solutions in fighting online misinformation. However, traditional fact-checking is a process requiring scarce expert human resources, and thus does not scale well to social media because of the continuous flow of new content \cite{hassan2015quest}. Automated methods and crowdsourcing have been proposed to tackle this challenge \cite{thorne2018automated,nakov2021automated,roitero2020can}, as they can scale with a smaller cost, but have always been studied in controlled environments. 
Current approaches focus either on fully automated machine learning methods \cite{wei-etal-2019-modeling,Liu2018EarlyDO} or on hybrid human-machine approaches making use of crowdsourcing to scale-up human annotation efforts \cite{10.1145/3340531.3412048}.

The first large-scale effort of crowdsourced fact-checking was piloted by Twitter with the \BW program on the 23rd of January 2021~\cite{BWPublish}. \BW adopts a community-driven approach for fact-checking by allowing selected Twitter users to identify fallacious information by (i) classifying tweets as misleading or not, accompanied by a written review, and by (ii) classifying reviews of other \BW users as being helpful or not.
In this setting, any user can create a \textit{note} for a tweet (providing some metadata about the annotations) and other users can up/down \textit{rate} such note. Multiple users can check the same content independently.

\begin{figure*}[t]
\includegraphics[scale=0.53]{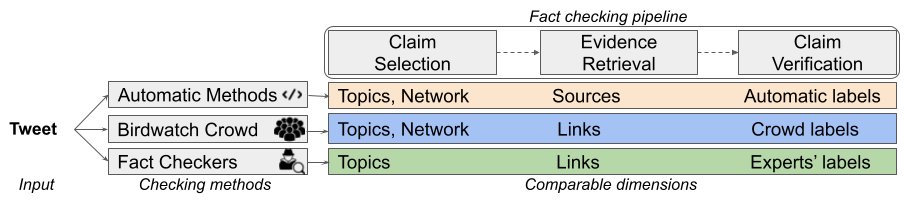}
\caption{Given input tweets, the three alternative checking methods \rev{(automatic, crowd, professional checkers)} are analyzed across their comparable dimensions according to a standard fact-checking pipeline.}
\label{fig:overview}
\end{figure*}

In this study, we perform an analysis of how crowdsourced fact-checking works in practice when compared with human experts and automated fact-checking methods.
To this end, we perform an analysis of the grass-root fact-checking process in \BW, including which content is selected to be fact-checked, which sources of evidence are used, and the fact-checking outcome. We also look at possible bias in terms of volume and topics as compared to experts.
To enable a fair comparison across the three fact-checking approaches (i.e., computational methods, crowd, experts), we collected a dataset of \rev{11.9k tweets with \BW checks and identified 2.2k tweets verified} both by \BW users and expert journalists. This dataset enables us to analyze and contrast the three approaches across the main dimensions in the standard fact-checking pipeline (see Figure \ref{fig:overview}). We focus on the following research questions:

\vspace{1ex}
\noindent \textbf{RQ1} How are check-worthy claims  selected by \BW users? Can 
the crowd identify check-worthy claims before experts do? 

\vspace{1ex}
\noindent \textbf{RQ2} What sources of information are used to support a fact-checking decision in \BW and how reliable are they? Does the crowd always rely on data previously fact-checked by experts, or can they be considered as ``independent fact-checkers''?

\vspace{1ex}
\noindent \textbf{RQ3} Are crowd workers able to reliably assess the veracity of a tweet? Is their assessment always considered helpful by others?

\vspace{1ex}
Our results reveal insights from real data to answer these questions. As automatic methods are still not competitive for checking the truthfulness of online content, we focus 
on how the crowd can fact-check claims and the way they do it as compared to experts. The main contribution of this work is 
an in-depth data-driven study of how crowdsourced fact-checking can work in practice, as compared to expert fact-checking over a number of dimensions such as topics, sources of evidence, timeliness, and effectiveness.

\vspace{1ex}
\textbf{Claim Selection.} The first step in the process is deciding which claims, out of the very many produced on Twitter, should be fact-checked. This is similar to the process of assessing relevance in a search task, as users are looking for a piece of content that is valuable and that satisfies their requirements (e.g., an information need or potential harm caused by the piece of content if misleading). Regarding the selection of the claims to check, we show that the crowd mostly matches the claims selected by expert fact-checkers in terms of topics, and it is not strongly influenced by properties of the social network, such as popularity of tweets. 
Moreover, we analyze the responsiveness of crowd and experts with respect to fresh tweets and found that in some cases the crowd is orders of magnitude faster in generating a correct fact-checking outcome.

\vspace{1ex}
\textbf{Evidence Retrieval.} In terms of sources, \BW users and fact-checkers rely on different set of online resources, with only few reference websites in common. For the  sources used by the crowd and the experts, we also compare the quality perceived by the \BW community against the quality ratings obtained  from a professional journalistic tool. The two scoring methods show correlation, but also remarkable bias in source quality assessment by the crowd on some topics related to politics.

\vspace{1ex}
\textbf{Claim Verification.}
In terms of effectiveness of the claim verification, we show that crowdsourcing may be an effective checking strategy in most settings, even comparable to the results obtained by human experts, but does not lead to consistent, actionable results for some topics. 
We also analyze the agreement among \BW users and the use of different scoring functions to aggregate their feedback, including the one used in production by Twitter.

\vspace{1ex}
Our observations show how crowdsourcing fact-checking in practice can bring an added value as compared to expert fact-checkers or computational methods used in isolation. Additionally, we release the first dataset of tweets with labels from expert fact-checker, crowd, and computational methods.

In the rest of the paper, we discuss related work in fact-checking and crowdsourcing (Section~\ref{sec:relatedWork}), introduce the datasets collected and crafted for our study (Section~\ref{sec:data}), and discuss the empirical results for our analysis (Section~\ref{sec:results}). Finally, we discuss the main challenges and opportunities for crowdsourced fact-checking 
(Section~\ref{sec:discussion}) and conclude the paper with some open research questions (Section~\ref{sec:conclusion}).

\section{Background and Related Work}
\label{sec:relatedWork}


Fact-checking requires a chain of steps that starts with identifying check-worthy claims and ends with a label about the veracity of the claim. Labels vary across services but usually can be divided into four popular categories: true, partially-true, false, or not enough evidence to judge. The top of Figure~\ref{fig:overview} shows a generic high-level fact-checking pipeline~\cite{nakov2021automated}. The three considered checking methods are then reported, specifically automatic methods, \BW crowd, and expert fact-checkers. Given an input textual tweet, every method can be used to assess if it is worth checking and eventually verified. For every checking method, we also report the dimensions that can be used to compare and contrast the alternative methods. 
We discuss next the main steps in the pipeline, \rev{their related work}, and their implementation in the different methods.

\textbf{Claim Selection.}
For claim selection we can use automatic methods, the crowd, or experts.
Given a sentence, an automatic method scores if it contains check-worthy factual claims~\cite{HassanALT17,clef-checkthat-T1:2019,10.1145/3308560.3316736}. A model trained on annotated sentences gives low scores to non-factual and subjective sentences.
Deciding whether a claim is worth checking is similar to the task of judging the relevance of a document w.r.t. a search query. In Information Retrieval evaluation, well-trained experts (e.g., NIST assessors) may be used to produce judgements of relevance following guidelines, or be instead substituted by crowd workers who receive simple instructions. 
Similarly, check-worthiness may be performed by a panel of experts 
or crowdsourced, like done in \BW.
The crowdsourced annotation of textual content on social networks is a widely supported activity across all platforms. Users label content that violates the guidelines of the site, such as hate speech and misinformation. This process triggers the human verification with moderators hired by the platform~\cite{FacebookProgram,10.1145/3404835.3464926}. For expert, human fact-checkers the selection of the claims to verify is driven by journalistic principles, e.g., claims should contain \textit{verifiable} facts~\cite{TruthOMeter}.
Experts also assess if a claim is \textit{important}, with a definition that changes according to the public and the mission of the organization, e.g., voters and elections~\cite{WaPost}.
The crowd may have different criteria and priorities in deciding which claims to fact-check and a  definition of check-worthiness that takes into account the topic, the timeliness, and their own personal points of view. Previous research in crowdsourced fact-checking (e.g., \cite{8791903}) has not looked in detail at how the crowd may perform this step of the pipeline, and it is something that instead we do in this work.

\begin{figure*}[t]
\includegraphics[scale=0.46]{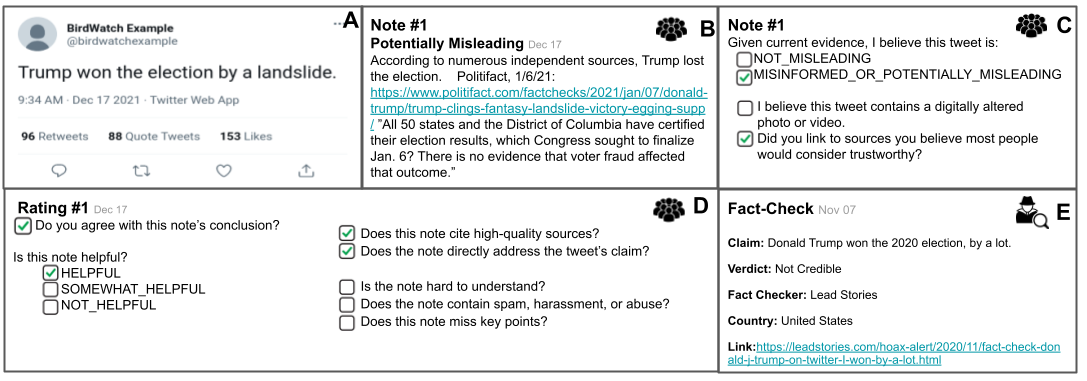}
\caption{\BW note and \CR fact-check Example. (A) shows a tweet. (B) is the note with the assigned label to such tweet. (C) is a sample of questions when submitting a note. (D) is a sample of questions when submitting a rating. (E) shows a fact-check delivered by an expert.}
\label{fig:bw_example}
\end{figure*}

\textbf{Evidence Retrieval.}
For computational methods, we distinguish the task of detecting previously fact-checked claims and the task of gathering evidence to support the verification step. As false claims are often repeated across platforms and over time, independently of  available fact-checks, claim matching aims at automatically identifying an existing debunking article for the claim at hand~\cite{adler2019,ShaarBMN20,botnevik2020brenda}. Claim matching is feasible at scale because websites use the schema.org standard \CR metadata
  to share their checks~\cite{CRProject}.
For fresh claims, which have not been debunked yet, several methods aim at finding external evidence to help  fact-checkers and computational methods deciding on the veracity of a claim~\cite{thorne-etal-2018-fact}. The output is usually a ranking of  retrieved documents or specific passages~\cite{Alshomary:20}.
The crowd makes use of expert fact-checking outcomes when available. Indeed, \citet{roitero2020can} removed expert outcomes from the search results used by the crowd in their fact-checking task to avoid influencing crowd worker judgments.
Expert fact-checkers instead rely on their training to identify proven, verified, transparent, and accountable evidence~\cite{evaluate}, sometimes involving third-party domain experts~\cite{fffaq}. 

\textbf{Claim verification.} A large body of research focus on developing and evaluating automatic solutions for fact-checking \cite{nakov2021automated,vo2018rise,BrandRSD21,botnevik2020brenda,10.1145/3404835.3463120,10.1145/3331184.3331305,Karagiannis0PT20}. However, there are coverage and quality issues with automated systems~\cite{fullfact:coof}, and thus a pragmatic approach is to build tools to facilitate human fact-checkers
~\cite{vo2018rise}.  At the same time, effort in artificially creating rumors and misinformation has been shown to be effective \cite{huynh2021argh}.
The crowd makes use of evidence from the Web and is influenced by their own personal belief and context \cite{10.1145/3340531.3412048,DBLP:conf/ecir/BarberaRDMS20}. Interestingly, when misinformation is identified on social media, users tend to counter it by providing evidence of it being misleading 
\cite{9377956}. This shows an intrinsic motivation that certain members of the crowd have to contribute to the checking process.
An approach for crowd-sourced fact-checking is using tools that surface relevant evidence for their judgement \cite{fan2020generating}. This however comes with the risk of over relying on such tools to make judgements  \cite{10.1145/3242587.3242666}.

Finally, there has been some early analysis of the \BW data~\cite{allen_martel_rand_2021,Prollochs2021CommunityBasedFO}, but they focus only on the tweets and notes content, while we rely on the manually aligned expert claim reviews to compare \BW results against the best solution in this space.
A related study has looked at a Reddit community involved in the fact-checking process using a crowdsourced approach \cite{10.1145/3308560.3316734}. 

\section{Data}
\label{sec:data}
Community-driven fact-checking on Twitter is governed by the \BW initiative~\cite{BWPublish}, while fact-checks written by journalists and expert fact-checkers are curated using the \CR schema~\cite{CRProject}. In this section, we describe both datasets and how to match similar claims identified by both parties. Approval from authors' institution research ethics committee to perform this study has been obtained prior to commencing.

\subsection{\BW}
\label{subsec:birdwatch}
Misinformation on Twitter can be mitigated through the \BW program, where participants can identify misleading tweets and provide more context~\cite{BWPublish}. Currently, \BW is only available to participants in the US, where users can identify misleading information using two core elements: \textit{Notes} and \textit{Ratings}.

{\textbf{Notes.}} Participants in the \BW program can add notes to any tweet. Their notes are formed from: (i) a classification label indicating whether the tweet is misinformed/misleading (\textbf{MM}) or not misleading (\textbf{NM}) according to their judgement, (ii) answers to several multiple-choice questions about their decision~\cite{BWdata},
and
(iii) an open text field where participants can justify their choice of the label and possibly include links to sources that prove their point. 
An example of a note is shown in Figure~\ref{fig:bw_example} (B,C). The key data we use from the notes are the following:
\begin{itemize}
    \item \textit{Classification Label}: Whether the tweet is 
    misinformed (MM) or not (NM) according to the \BW user (Section~\ref{subsec:ClaimVerif}).
    \item \textit{Note Text}: the text given by the user with the justification for the label (Sections ~\ref{subsec:topics} and \ref{subsec:rq1}).
    \item \textit{Timestamps}: time at which the note was written \rev{(Section~\ref{subsec:rq1}}).
\end{itemize}

{\textbf{Ratings.}} Participants rate the notes of other participants. Ratings help identify which notes are most helpful.
A user rates a note by providing answers to a list of questions~\cite{BWdata}. An example of a rating is shown in Figure~\ref{fig:bw_example} (D).
Out of these questions, we focus on the following:
\begin{itemize}
    \item \textit{High-quality Sources:} The user answers the yes/no question `Is this note helpful because it cites high-quality sources?'. We use this information to assess whether \BW users distinguish credible sources~(Section \ref{sec:sourceQuality}).
    \item \textit{Helpfulness Label:} The user answers the question `Is this note helpful?'. The possible answers are (i) not helpful, (ii) somewhat helpful, and (iii) helpful. We use this information to compute an helpfulness score for notes (Section~\ref{subsec:ClaimVerif}).
\end{itemize}

All \BW notes start with a `Needs More Rating' status until enough ratings are achieved according to a platform defined threshold (currently set to 5). Once achieved, these ratings are aggregated and weighted by a `Rater Score' to compute the `Note Helpfulness Score'. A higher rater score gives more weight to participants (i) whose notes are found helpful by other participants, and (ii) whose ratings align with the final rating outcome. A higher note helpfulness score means that many participants found a note adequate, and it would likely hold a valid classification label. 

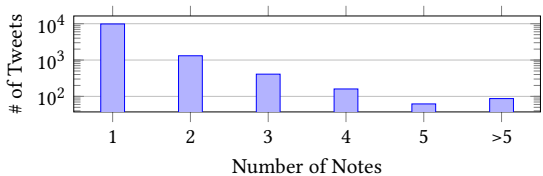
\begin{figure}[h]
    \centering
    \pgfplotstableread[col sep=comma,]{files/Notes.csv}\datatable
    \begin{tikzpicture}[scale=0.9]
    \begin{axis}[
    	ymode=log,
	log basis y={10},
        ybar,
        xlabel={Number of Notes},
        xtick=data,
        xticklabels from table={\datatable}{NumNotes},
        ylabel={\# of Tweets},
        ymajorgrids,
        height=3cm,
        width=\columnwidth]
        \addplot table [x expr=\coordindex, y={Count}]{\datatable};
    \end{axis}
    \end{tikzpicture} 
    \caption{Bar plot of the number of notes per tweet.}
    \label{fig:BarNotes}

\end{figure}
\begin{figure}[h]
    \centering
    \pgfplotstableread[col sep=comma,]{files/NumRate.csv}\datatable
    \begin{tikzpicture}[scale=0.9]
    \begin{axis}[
        ybar,
        xlabel={Number of Ratings},
        xtick=data,
        xticklabels from table={\datatable}{NumRating},
        ylabel={\# of Notes},
        ymajorgrids,
        height=3cm,
        width=\columnwidth]
        \addplot table [x expr=\coordindex, y={Count}]{\datatable};
    \end{axis}
    \end{tikzpicture}
    \caption{Bar plot of the number of ratings per note.}
    \label{fig:BarRates}
\end{figure}
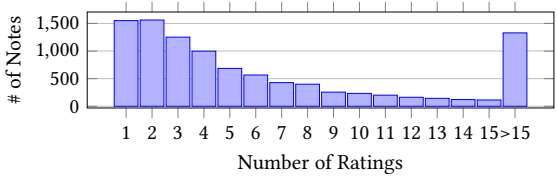

{\textbf{Descriptive Statistics.}} We use the \BW data up to September 18$^{th}$ 2021. The dataset contains 86,924 ratings for 15,445 notes on 11,871 tweets from 5124 unique \BW participants. Bar plots of the number of notes and ratings are shown in Figure \ref{fig:BarNotes} and Figure~\ref{fig:BarRates}, respectively. Most tweets have only one or two notes, while
the tweet with the most notes has 61. The majority of notes have less than five ratings, and the most rated note has 184 ratings. The user with most notes checked 656 tweets with around 71\% related to US Politics. 
Among these 656 tweets, 643 do not have any other note. The user with most notes in common with other users shares 85 notes (on 85 tweets) with 217 other users. 

\subsection{\CR}
\label{subsec:ClaimReview}
The \CR project~\cite{CRProject} is a schema used to publish fact-checking articles by organizations and journalists. The schema defines 
mark-up tags that are used in web pages so that search engines identify the  information in a debunking article, such as text claim, claim label, and author~\cite{CRSchema}.
Our dataset is a collection of items following the \CR schema, collected from various sources 
\cite{MensioA19}. Each item, or \textit{fact-check}, is a (claim, label) pair produced by a professional journalist or fact-checking agency.
\rev{We assume that professional fact-checkers do not overlap with \BW participants, as the former have no interest in doing their work without retribution.}
Since different fact-checkers use different labels, the data is normalized into a smaller subset of labels (credible, mostly credible, uncertain, unverifiable, not credible). In addition to the claim and the label, the checks also contain a link to the fact-checking article. 
Note that checked claims in this dataset could occur anywhere on the web and need not be only on Twitter.
We use a dataset containing 76,769 fact-checks. 
Examples of the data are shown in Figure~\ref{fig:bw_example} (E) and Table~\ref{tab:examples}.

\subsection{Matched Data}
\label{subsec: MatchedData}
To study how the judgements of the crowd compare to those of expert fact-checkers, we match claims from both datasets. 
\rev{As the automatic matching is imperfect, we used the Amazon Mechanical Turk crowdsourcing platform~\cite{AMTurk} for matching the text in the tweets checked by the \BW crowd with the claim text in the \CR fact-checks. When workers accepted a Human Intelligence Task (HIT), they were shown (i) the tweet that is to be matched and (ii) the top-10 similar \CR checks provided by SentenceBERT using a bi-encoder with cosine similarity \cam{between the text of the tweet and that of the claim in \CR fact-checks}~\cite{reimers-gurevych-2019-sentence}. We also add a `None of the Above' option for cases where the worker could not find a match.
A manual inspection of the matches showed that the vast majority of tweets with a score below 0.6 do not have matching \CR checks. We therefore run the annotation for tweets with at least 0.6 as top-1 similarity score. The workers were required to have at least 500 approved HITs to access our task, which comprised of 5322 tweets to be matched. Each tweet was shown to 3 workers\cam{, similar to previous work~\cite{Kazai2011,Du2012}.} 
\cam{The hourly rate based on median completion time was 12.41\$.}
}


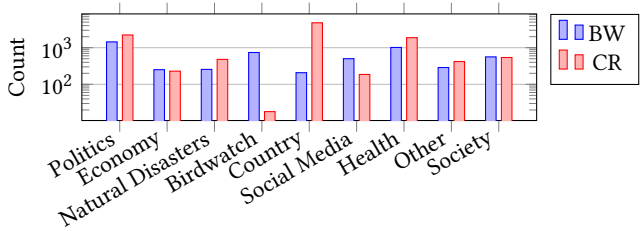
\begin{figure}[t]
\pgfplotstableread[col sep=comma,]{files/TopicBar.csv}\datatable
\begin{tikzpicture}
\begin{axis}[
	ymode=log,
	log basis y={10},
    height=3cm,
    width=0.9\columnwidth,
    ymajorgrids,
	ylabel=Count,
	legend pos=outer north east,
	enlargelimits=0.1,
	ybar,
	bar width=4pt,
	xtick=data,
	xticklabels from table={\datatable}{Topic},
	x tick label style={rotate=30,anchor=east}],
]
\addplot
     table[x expr=\coordindex,y=BirdWatch]{\datatable};
\addplot
     table[x expr=\coordindex,y=Claim Review]{\datatable};

\legend{BW,CR}
\end{axis}
\end{tikzpicture}
\caption{Bar plot of tweets checked by \BW (BW) and \CR (CR) fact-checks for 2021 divided by topic.}
\label{fig:2021_topics}
\end{figure}

\begin{figure*}[t]
\includegraphics[scale=0.35]{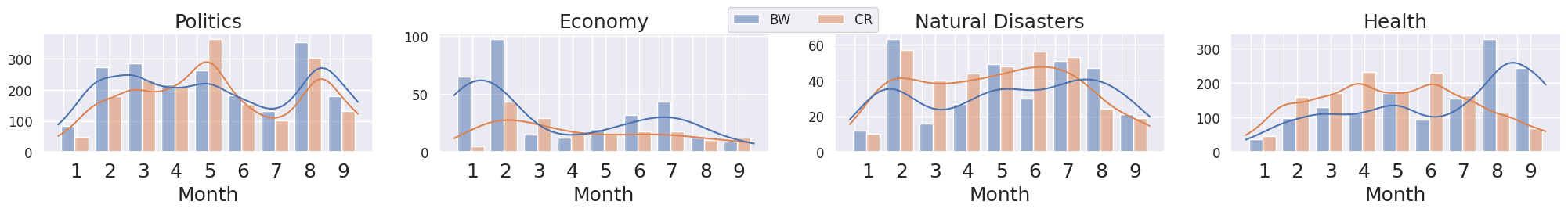}
\caption{Per-topic frequency histograms and KDE Plots for \BW (BW) notes and \CR (CR) fact-checks (month granularity).}
\label{fig:TopicPlot}
\end{figure*}

\rev{To measure the quality of the worker annotations, we manually annotated the top-500 tweets in terms of matching score. 
Among these 500 tweets, we manually identified 75 with a matching \CR check. Workers correctly matched 63/75 tweets 
according to our ground truth, while the baseline method choosing the highest SentenceBERT score correctly matched 59/75 tweets.
After running the study over the 5322 tweets, we obtain 2208 tweets (3043 notes) matching with \CR checks. 
An example of a tweet matching a \CR check is shown in Figure~\ref{fig:bw_example} (A,E). More examples of matched tweets, \BW notes, and \CR checks are in Table~\ref{tab:examples}. Our dataset containing matched tweets to \CR checks alongside labels from \BW and \CR \cam{ and code relevant to the paper are} available at \url{https://github.com/MhmdSaiid/BirdWatch}.}

\subsection{Topics}
\label{subsec:topics}
We analyze how \BW notes \rev{and \CR checks compare in terms of covered topics}. We use \textit{BertTopic}, a topic-modeling technique that utilizes transformers and TF-IDF for clustering~\cite{grootendorst2020bertopic}, 
to predict the topic of every \BW tweet and \CR \rev{claim} for the year 2021 and report their frequency distributions in Figure ~\ref{fig:2021_topics}.
\textit{Politics} and \textit{Health} have high counts in both. Topic \textit{Country}, which includes news about countries all over the world, has higher counts for \CR data since \BW is deployed in the US only. \BW notes cover mostly tweets in English and is biased towards US related news, whereas the \CR data contains fact-checks in different languages and from local fact-checking agency, thus explaining the high number of country-related tweets. 






\section{Results}
\label{sec:results}
We report results in addressing our three research questions next.

\subsection{RQ1: Claim Selection}
\label{subsec:rq1}
We analyze how \BW participants effectively identify check-worthy claims in a comparison with fact-checking experts.
We also compare \BW users, who do not necessarily have journalistic training, against computational methods for this task.

\subsubsection{Topic Analysis} After predicting the topic of every \BW tweet and \CR fact-check, we plot the frequency distribution of four topics showing interesting trends, on a monthly basis, in Figure~\ref{fig:TopicPlot}. The high count of \BW tweets and \CR fact-checks covering political tweets show that they both consider the \textit{Politics} topic important. The similar trends for this topic suggest that both methods react similarly to news and major events in terms of claim selection. For example, 
the peak in \textit{Politics} for both methods in August is related to the Taliban take-over of Afghanistan. 
We observe the same trend for the topics \textit{Economy} and \textit{Natural Disasters}. 

However, for the  \textit{Health} related tweets, we observe an abrupt change in the trends from July 2021. This is due to the emergence of the COVID-19 Delta variant in US, which triggered more tweets about the topic, mainly discussing masks/vaccines issues, and more \BW notes on this topic. 
This is accompanied by a decrease in the number of health-related fact-checks, which can be explained by multiple reasons. 
One explanation is that the most important issues about masks and vaccines had already been debunked before the Delta variant. This shows that \rev{despite} fact-checks are available online, numerous social network users keep spreading false claims that have previously been debunked (see also Section \ref{subsec:ClaimVerif}). 

Topic selection also reflects the different geographic focus of the two methods. For example, the \BW peak in February in topic \textit{Economy} is due to the Texas power crisis, 
a US-specific event.
\rev{Despite the differences, our results show that both \BW participants and \CR experts pick the content to verify in response to the events happening in reality, independently from the specific topic.}

\begin{figure}[t]
    \centering
    \includegraphics[scale=0.44]{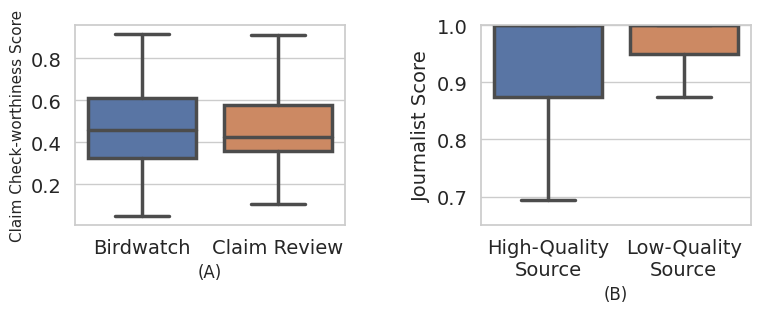}
    \caption{(A) shows a box plot of claim check-worthiness scores of \BW tweets and \rev{the claims in the} \CR fact-checks. (B) shows a box plot of journalist scores compared to the final verdict of \BW users \rev{(x-axis)}.}
    \label{fig:BoxPlots}
\end{figure}

\subsubsection{Computational Methods} We report on the ClaimBuster API for claim check-worthiness~\cite{HassanALT17}. Given a sentence, the API provides a score between 0.0 and 1.0, where a higher score indicates that the sentence contains check-worthy claims. We run the API on \BW tweets and \rev{the claims in the} \CR fact-checks, with the associated box plots for the scores in Figure~\ref{fig:BoxPlots} (A). \rev{The results show a check-worthiness median score at around 0.4, for both sets of claims, while in ClaimBuster the suggested threshold for check-worthiness is 0.5~\cite{HassanALT17}. One explanation of the difficulties of computational methods for claim selection is the bias in the training data used to build them. Indeed, most available datasets for this task are of high-quality text, coming from articles or political speeches, while the text used on Twitter is usually much noisier, e.g., due to the use of slang}.

\begin{figure}
    \centering
    \includegraphics[scale=0.41]{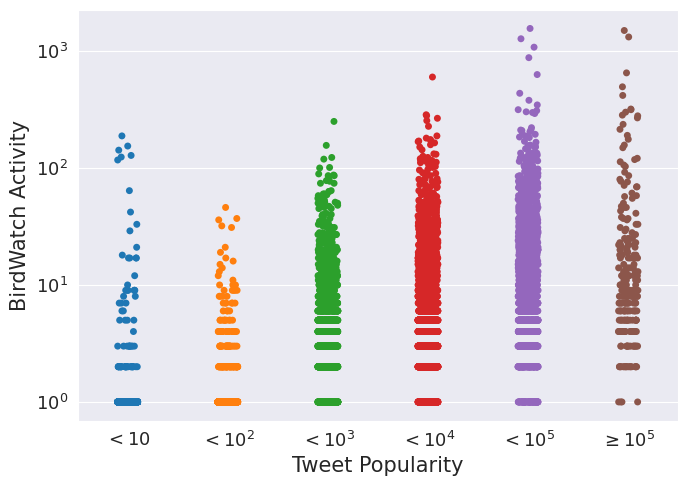}
    \caption{Tweet popularity and \BW activity.}
    \label{fig:pop}
\end{figure}

\subsubsection{Tweet Popularity}
We check whether the claim selection process of \BW users is affected by the popularity of a tweet. For every tweet, we retrieve the number of retweets and favorites and sum them to obtain a quantifiable popularity score. As expected, Figure~\ref{fig:pop} shows that popular tweets receive more activity than others from the \BW community, i.e., have more notes and ratings. However, there are popular tweets with low \BW activity and unpopular tweets with high number of notes and ratings. 


\subsubsection{Temporal Analysis} We analyze tweets (T), \BW notes (B), and \CR fact-checks (C) time-wise. As a note can only occur after a tweet, we have three different configurations: (i) Tweet occurs first, then \BW note, then \CR fact-check (TBC), (ii) Tweet then \CR fact-check then \BW note (TCB), and (iii) \CR fact then Tweet then \BW note (CTB).

\textbf{TBC:} There are \rev{129/2208} tweets in our matched data for this case. In all tweets, \BW users provide a response much faster than experts. On average, a \BW provides a response 10X faster than an expert. These examples show how \BW participants can fact-check claims with reliable sources without the need of \CR fact-checks such as ID \#4 in Table~\ref{tab:examples}.

\textbf{TCB:} In our dataset, a \CR rarely occurs after a tweet and before a \BW. We observe faster responses from \CR than \BW users for \rev{26/2208} tweets.  Since the granularity of the \CR is days while that of \BW is seconds, there are also \rev{17/2208} tweets that occur on the same day, and we cannot state which of the two was actually faster. 

\textbf{CTB:} The majority of the matched tweets follow this pattern, with most of them related to US politics and COVID-19. As Twitter is an open space, several users tend to spread false news even after they have been fact-checked, specifically those related to Trump winning the elections. We discuss more this issue in Section~\ref{sec:discussion}.

\vspace{1ex}
{\emph{Claim Selection Take-away Message:}} \BW users and \CR experts show correlation in claim selection decisions w.r.t. major news and events, but with important differences due to the circulation of claims that have been already debunked by experts.  The crowd seems to be effective also in identifying tweets with misleading claims even before they get fact-checked by an expert.
Also, both popular and non-popular tweets get verified by \BW users.
\rev{Computing} the check-worthiness of a tweet does not lead to effective results using current \cam{off the shelf APIs}.

\subsection{RQ2: Evidence Retrieval}
Both crowd checkers and experts report the sources used in their verification process.
We analyze such sources and then contrast their quality according to an external journalistic tool. 

\subsubsection{Descriptive Statistics}

We extract all links from \BW notes. We find a total of 12,909 links covering  2,014 domains. Unsurprisingly, the top cited links are those coming from journalistic and fact-checking sites (Politifact, Reuters, NYtimes) and governmental websites, such as USGS and CDC. The distribution of the links is right-skewed, where half of the links are from only 29 domains.
\CR checks contain 76,769 links covering only 73 domains of fact-checking groups and journalists. The distribution of links shows less skewness than that of \BW.

\BW participants use only 17 domains in common to those of the \CR experts. The other 56 \CR domains, which are not in the overlap, include 53 
local resources, such as news outlets, for non US countries as fact-checking organizations work at a global scale and \BW focuses on US.  
\BW sources are a larger number as they range from Wikipedia and YouTube videos to medical websites and research papers.

\subsubsection{Expert Judgement of Source Quality}
\label{sec:sourceQuality}
We compare ratings of source quality of \BW users to those of expert fact-checkers.
To assess the quality of web sources, we rely on an external tool that provides a score (between 0 and 100) where the higher the score, the higher the quality of the source\footnote{\url{https://www.newsguardtech.com}}. The score is obtained by journalists manually reviewing every website, and we refer to it as the \textit{journalist score}.

For every note in our matched dataset, we first compute a \BW score indicating whether the links are high-quality sources or not by performing a majority voting on the ratings of the note, and we then compute the average journalist score of every link in the note. \cam{Out of 3043 notes, 2231 contained links. We obtained results for \rev{656}, while the others either had (i) no ratings (363/2231), (ii) no journalist scores (698/2231), (iii) nor both (309/2231), or (iv) there was no majority in the ranking votes (205/2231).
}


A box plot of journalist scores and \BW labels is shown in Figure~\ref{fig:BoxPlots} (B). For note links rated as high-quality by \BW users (with majority voting), we observe high journalist scores. The majority of tweets of the notes are related to US elections \rev{and COVID-19}, with \BW users citing sources such as Politifact and \rev{CDC}. \rev{Some sources in the notes have been classified as being high-quality by \BW users but low-quality w.r.t. the journalist scores. Those notes share mainly COVID-19 studies such as Mayoclinic.org, a nonprofit American medical center, and fda.gov, the US food and drug administration, that are regarded as reliable sources in the US but do not meet all the requirements for high journalist score.}


\cam{238/656 notes contain sources that are \rev{rated as low-quality}, but have a high journalist score.} These notes are debunking news about US politics, specifically about Trump winning the 2020 elections \rev{and misinformed COVID-19 content}. These tweets include links to reliable sources, but a significant fraction of \BW users labeled such links as low-quality. This shows how some \BW users convey partisanship, forming a group of people trying to deceive the \BW program to serve their common interest, such as supporting a political party in social media.

Such groups can be effective in ``gaming'' the algorithm~\cite{EpsteinPR20}, ultimately having a profound effect on \BW since (i) the biased \BW participants can steer the ultimate label of a note to their favor, thus spreading misinformation, and (ii) by increasing their weight in the \BW platform since if one's ratings match those of the ultimate rating, they will get a higher weight. \cam{As an example, for the tweet \texttt{`Joe Biden is President In Name Only. \#PINO'}, a certain note replied that Biden is indeed the president with links from \textit{Politifact} and \textit{APNews}, both having journalistic scores of 100/100 and 95/100 respectively while 12/14 of the raters identified such sources as unreliable. }

We also compute journalist scores for links in \BW and \CR data. As \BW users use many links, we only computed scores for the top-100 occurring links that form 68.6\% of the data. While both distributions of \BW and \CR link scores attain a median of 1.0, links by \CR fact-checks have lower variance with a minimum of 0.875, while that of \BW notes is 0.495.

\vspace{1ex}{\emph{Evidence Retrieval Take-away Message:}} Expert fact-checkers rely on a relatively small set of high-quality sources to verify claims, while \BW participants provide a variety of sources that seem to be neglected by fact-checkers. While most of these sources are evaluated as credible (by journalists) and useful (by the \BW crowd), malicious users might game the algorithm and effectively label notes as unhelpful according to their ideology. 

\begin{figure}[t]
\includegraphics[scale=0.35]{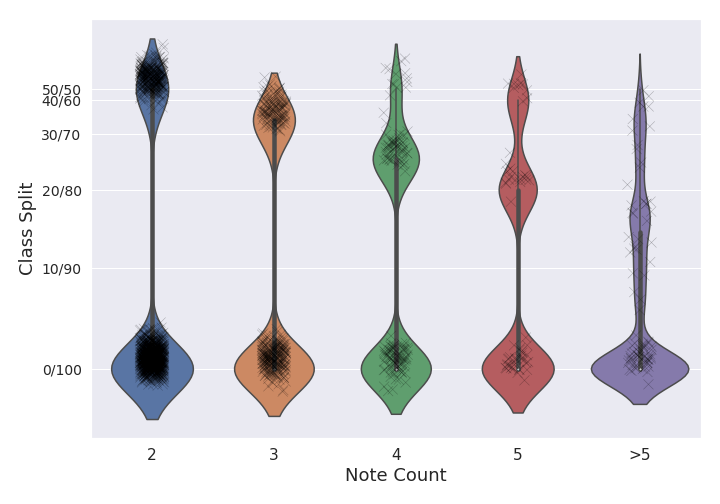}
\caption{Violin Plot of note counts and class splits of the classification labels of notes. The figure shows the Kernel Density Estimation (KDE) plot of every note count, with their respective data points. Tweets with two notes are dominant, with most notes agreeing on the label. Other note counts do show full agreement on the label (0/100), with more cases of disagreement as the number of notes increase.}
\label{fig:ViolinPlotClass}
\end{figure}

\subsection{RQ3: Claim Verification}
\label{subsec:ClaimVerif}
We ponder whether \BW participants provide accurate judgements. We first compare agreement (i) among themselves and then (ii) with \CR expert fact-checkers. We then analyze different scoring functions for note aggregation, and finally \rev{report results for computational methods.}

\subsubsection{Internal Agreement}
\label{subsubsec:internal}
We use the participants' classification labels to see whether the tweet is classified as misinformed or not.
To compute agreement, we use the standard metrics Krippendorff’s alpha~\cite{Krippendorff2011ComputingKA} and Fleiss's kappa~\cite{Fleiss1971MeasuringNS}. However, due to the large sparsity in the data and the huge number of missing values, both metrics fail to provide meaningful results \cite{checco2017let}. We then compute the variance as a metric for agreement. Lower variance means that all \BW participants agree on the classification label.  

A violin plot is shown in Figure~\ref{fig:ViolinPlotClass} for tweets with various note counts.
On the y-axis, we report the density of the class splits, where a class 0/100 indicates full agreement across the users and 50/50 indicates full disagreement.
We see that most tweets have two notes and the majority of users perfectly agree on the final classification label. The same applies to tweets with more note counts, where most of the notes agree on the final label, with conflicts happening on some tweets but with a small subset with full disagreement. A topic analysis of tweets shows that 48.3\% of tweets with full disagreement are related to either politics or COVID-19. 


\begin{table}
\setlength{\tabcolsep}{.445em}
\begin{tabular}{cccclccc}
\toprule
\multicolumn{1}{l}{}                                                             &
\multicolumn{1}{l}{}          & \multicolumn{5}{c}{\textbf{BirdWatch}}                                                          \\
\midrule
\multicolumn{1}{l}{\textbf{}}                                                    & \textbf{}                     & \multicolumn{2}{c}{\textbf{Notes}}            &           & \multicolumn{3}{c}{\textbf{Tweets}} \\
\midrule
\multicolumn{1}{l}{}                                                             & \multicolumn{1}{l}{\textbf{}} & \multicolumn{1}{l}{\textbf{MM}} & \textbf{NM} & \textbf{} & \textbf{MM}      & \textbf{NM}  &\textbf{Tie}    \\
\multirow{5}{*}{\textbf{\begin{tabular}[c]{@{}c@{}}Claim\\ Review\end{tabular}}} & \textbf{credible}             & 209                               & \textbf{25}           &           & 126                & \textbf{9}    &9            \\

                                                                                 & \textbf{mostly\_credible}        & 56                             & \textbf{14}           &           & 44              & \textbf{7}  &5              \\
                                                                                 
                                                                                 & \textbf{not\_credible}        & \textbf{1983}                             & 184           &           & \textbf{1476}              & 62  & 55              \\
                                                                                 
                                                                                 & \textbf{not\_verifiable}      & 300                              & 25           &           & 225               & 8   &9             \\
                                                                                 
                                                                                 & \textbf{uncertain}            & 225                               & 22           &           & 156                & 8  &9      \\
\bottomrule
\end{tabular}
\caption{Matching the classification labels across \BW and \CR on the note level and the tweet level (obtained through majority voting). Agreement in bold.}
\label{tab:extAgr}
\end{table}

\setttsize{\Small}

\begin{table*}[t]
\small
\resizebox{0.99\textwidth}{0.255\textheight}{
\begin{tabular}{lp{3.1cm} p{4.2cm}cp{3.5cm}l p{3.9cm}}
\hline
\textbf{ID}  & \textbf{Tweet}      & \multicolumn{2}{c}{\textbf{BW Note}}               & \multicolumn{2}{c}{\textbf{CR Fact}} & \textbf{Comment}   \\ \cline{3-6}
    &                      & \multicolumn{1}{l}{\textbf{Text}} & \multicolumn{1}{l}{\textbf{Label}} & \textbf{Text}    & \textbf{Label}    &                    \\ \hline
\multicolumn{1}{c}{\textbf{\#1}} & \texttt{Pregnant women, please don't take this vaccine. \url{https://t.co/4KKlnMIbl7}} &\texttt{Updated CDC guidance, and newly accepted and reviewed medical research, has stated there are no safety concerns for pregnant women to be vaccinated against COVID-19.} \emph{(links omitted for brevity)}
& MM                             & \texttt{The vaccine is not safe for pregnant women or women planning on becoming pregnant within a few months of taking the vaccine... We are the lab rats.}        & NC          & The \BW user provides proof of why the claim is False. The fetched fact-check has a label of being not credible. \\
\midrule

\textbf{\#2}&\texttt{The mass shooting at Marjory Stoneman Douglas High School in Parkland, Florida was real and not staged.}&\texttt{That this continues to be debated is astounding.  Yes, this really happened.  Here is a link:    \url{https://en.m.wikipedia.org/wiki/Stoneman\_Douglas\_High\_School\_shooting}}&NM& \texttt{Say David Hogg is a crisis actor.}&NC&BW note confirms the tweet, thus the label was not misleading. The CR check states a claim opposite to the tweet and its label is not credible.\\
\midrule
\textbf{\#3}&\texttt{Chicago PD Says Enhanced Vid Shows Gun in 13-Year-Old Adam Toledo's Hand \url{https://t.co/B0Twu733RL}}&\texttt{Chicago Mayor Lori Lightfoot said Adam Toledo had a gun in his hand when he was fatally shot by a police officer, or words to that effect.}&MM&\texttt{Adam Toledo did not have the gun in his hand when he was approached by the police who shot him. He has his arms up and complied. The gun was on the floor, not his hands.}&CR&Difference in granularity of the claim. For CR check, the claim is whether Toledo was holding a gun; while for the BW note, the claim was whether Chicago's mayor \textit{said} that Toledo hold a gun.\\
\midrule
\textbf{\#4}& \texttt{New poll indicates Biden approval at 11\%. The LOWEST approval rating of ANY president in American history. ~Gallup via Daily Caller}&\url{https://fivethirtyeight.com/features/how-were-tracking-joe-bidens-approval-rating/}&MM&\texttt{Biden approval at 11\%}&NC&The \BW participant provided a reliable source (score of 92.5) 7 days before a fact-check by an expert was available.\\
\midrule
\textbf{\#5}& \texttt{Biden thinks he came to the US Senate 120 years ago?!?}&\texttt{US President Joe Biden made a clear joke at his first press briefing since his inauguration, in which he said he went to the Senate 120 years ago. This is a self-deprecating joke and shouldn't be taken seriously.} 
&MM&\texttt{Joe Biden said, `With regard to the filibuster, I believe we should go back to the position of the filibuster that existed just when I came to the United States Senate 120 years ago.'}&CR&The Tweeter took a joke seriously, which was interpreted as misleading by the \BW participant.\\
\bottomrule
\end{tabular}}
\caption{Examples of Tweets, \BW Notes, and matched \CR fact-checks. \BW uses labels misleading/potentially misinformed (MM) or not misleading (NM), while \CR uses credible (CR) or not credible (NC).}
\label{tab:examples}
\end{table*}

\subsubsection{External Agreement}
\label{subsubsec:external}
After matching \BW data with \CR fact-checks, we compare their labels. Table~\ref{tab:extAgr} shows that the majority of \CR labels match the \BW ones. Specifically, in terms of notes, there are \rev{2022 cases (25+14+1983)} where they agree and \rev{449 cases} of disagreement. In terms of tweets, there are \rev{1492 (9+7+1476) decisions with the same classification label and 232 (126+44+62) with different labels. For 69 (9+5+55) tweets there is a tie in the voting across \BW users.} For completeness, we report also the numbers for other \CR labels (`not\_verifiable' and `uncertain'), even if they have no  mapping to \BW classification labels. 
\rev{We did some analysis to understand the cases where the labels are not aligned, some examples are reported in Table~\ref{tab:examples}.
Among the 209 notes that are labeled as credible by the \CR fact-checks and misinformed by the \BW participants, the most common cause are texts with multiple claims, 
i.e., multiple facts are reported in a tweet 
and the fact-checked claims differ (ID \#3). In other cases, tweets are mistakenly labelled as misinformed, e.g., because a joke is taken seriously by a Twitter user (ID \#5). Finally, assuming correct \CR labels, we believe in some cases the mismatch is due to biased \BW users. For the tweets labeled as not credible by \CR fact-checks and not misleading by \BW notes, we observe cases where a \BW note is the negated version of the \CR fact-check (ID \#2), thus producing opposite labels. There are also mismatch of labels, even though the \BW user provides evidence from a link that has a high journalistic score (0.875).}

\subsubsection{Note Helpfulness Score}
In the real-world production setting, not all \BW notes are used for finding the ultimate label that gets exposed on the platform. In fact, a \textit{note helpfulness score} is computed by the platform for each note, and those having a high enough score are used for computing the ultimate label. \BW exposes the code for computing such score, however, the public code does not include raters' scores into consideration. We use the available code and filter out notes that are not helpful for the final label, using a Twitter-defined threshold for the note helpfulness score (0.84). We are left with \rev{533 tweets (over 2208)} that pass the threshold, \rev{with 333 notes labeling the tweets according to \CR checks. About 95\% of notes label the tweets as misleading, thus} 
indicating that \BW users tend to rate misleading tweets more than non-misleading ones, in agreement with previous work~\cite{Prollochs2021CommunityBasedFO}. Of course, malicious ratings of the classification labels can steer the note helpfulness score in misleading directions, similarly to the judgement for source quality as discussed in Section~\ref{sec:sourceQuality}.

\subsubsection{Computational Methods.} We compare our matched data with labels coming from computational fact-checking systems. We use again ClaimBuster, \rev{as it can also verify claims}~\cite{HassanZACJGHJKN17}, and E-BART~\cite{BrandRSD21}.
ClaimBuster provides correct results for \rev{118 out of 2208 tweets, where 2090 tweets have no output from the model with an F1-score of 0.042. E-BART correctly labels 369 (over 2208) and does not produce a decision for 59 tweets with an F1-score of 0.17. A random classifier produces an average F1-score of 0.333 with 0.008 standard deviation.} As for claim selection, tweets are harder to handle for computational methods than news articles and quotes from politicians, which are the bulk of content in training corpora.

\vspace{1ex}
{\emph{Claim Verification Take-away Message}}: \BW users show high enough levels of agreement to reach decisions in the vast majority of cases. The \BW crowd focuses mostly on misleading tweets and shows high agreement with expert fact-checkers in terms of classification label. 
Computational methods have room for improvement in automatically verifying tweets.

\section{Discussion}
\label{sec:discussion}

\subsection{Collaborative solutions}
The analysis of the quality of the \BW users 
shows that crowdsourced fact-checking is a promising and complementary solution, with results that correlate with those of professional fact-checkers. However, we argue that a crowd-based solution should not be considered to replace experts, but rather as a tool in collaborative effort where, for example, the crowd helps flagging content and creating links to more sources of trustable evidence. Indeed, our results show that the crowd can be even more reactive than experts to a new false claim and is able to identify a large array of high quality sources of evidence. This is especially important, as there is evidence that fact-checking interventions are significantly more effective in novel news situations~\cite{Nevo2022}.

Looking forward, and assuming we can characterize the trust and the cost level for all involved actors (i.e., crowd, experts, and computational methods), there is an opportunity to design novel hybrid human-machine solutions that coordinate this joint effort in order to combine the benefits of the different approaches. The role of automatic tools can be that of providing real-time and scalable fact-checking for all posted content. Platform users can then intervene quickly in a more focused manner to provide a first line of defense on potentially harmful content. This can then be followed by quality in every step of the fact-checking pipeline, with humans collectively processing evidence for the final labeling. 

\subsection{Hard to verify claims}
The matching process of tweets and claim-review checks led us to recognize the difficulty of this task. 
The first challenge is 
the semantic match in terms of content, but in many cases where the match is clear, the problem is hard even for humans. Several problems, such as sarcasm and vagueness, are known in general for the detection of worth-checking claims~\cite{clef-checkthat-T1:2019}. However, another problem is the granularity of the tweet. Even a very short tweet may contain two interleaved claims, such as ``Mike said: the earth is flat'' (see also, e.g., Tweet \#3 in Table \ref{tab:examples}). Assume there are two claim reviews, one checks that Mike made a claim about the earth (labeling the matching tweet as true), and the other checks about the fact that the planet is not flat (labeling the tweet as false).

This suggests the challenge of being able to identify the textual claims where both crowdsourcing and computational fact-checking methods are most likely to fail short. This can be modeled as a new supervised classification task aiming at predicting when a claim cannot be verified effectively without experts. 
{As an orthogonal approach, this is also an opportunity for automatic controversy detection methods (e.g., \cite{10.1145/2911451.2914745}) to play a complementary role in supporting the crowd, making them aware of potential controversies during their verification tasks.} 

\subsection{Stale claims}
We found clear evidence that claims that have been already verified by expert checkers keep circulating and spreading on Twitter, even months after the publication of their debunking~\cite{shin2017}.
Unfortunately, we have also observed that automatic methods still fail short in matching with high accuracy tweets that contain such ``stale" claims{, likely because of the peculiar language used in tweets.}
In such a setting, 
\BW users can play an important role in quickly and effectively recognizing these cases.
Indeed, the significant difference in the health-related \BW notes and \CR fact-checks is explained by the increase of tweets spreading already fact-checked claims.
Stale claims are a good fit for the role of \BW participants, especially when automatic matching methods~\cite{ShaarBMN20,AhmadICDE} fail, while fresh claims might require proper forensic processes and need the expertise of journalists. 

\section{Conclusion}
\label{sec:conclusion}
In this paper, we present a data-driven analysis of the \BW program through the lens of the three main components of a fact-checking pipeline: claim detection based on check-worthiness, evidence retrieval, and claim verification. This is also the first study that bridges real data from a large-scale crowdsourced fact-checking initiative with the debunking articles produced by professional fact-checkers. \cam{While our analysis has been limited to \BW participants from the US, we hope the \BW initiative can be deployed globally for a more comprehensive analysis.}

Our study shows that 
\BW notes are effective in terms of claim verification, with encouraging results, in contrast with the negative results obtained by previous crowdsourcing efforts
~\cite{BhuiyanZSM20}. However, we also show that in \BW a group of users sharing a common goal could potentially steer the final classification label, such as in the case of source credibility. This suggests that more attention is needed in identifying harmful groups by profiling their activity and by incorporating their biases in the note ranking system. Another approach would be to calibrate the selection of \BW participants to enforce high diversity 
to mitigate this issue~\cite{EpsteinPR20}.

Our results indicate  that the \BW program is a viable initial approach towards crowd-based fact-checking, which can help complement the work of expert fact-checkers.
An interesting open question is how to develop a collaborative platform involving the different fact-checking methods (crowd-based, computational, and experts). 
Given different trust and cost profiles for the different methods, a model to assess the difficulty of validating a claim (either data-driven  or crowd-driven as in \BW), and a certain budget, what is the optimal way to assign the claim to each fact-checking method so that the number of verified claims is maximized with a high level of trust?



\paragraph{Acknowledgments.} This work is partially supported by an ARC Discovery Project (Grant No. DP190102141), by the ARC Training Centre for Information Resilience (Grant No. IC200100022), by gifts from Google and by CHIST-ERA within the CIMPLE project (CHIST-ERA-19-XAI-003).

\bibliographystyle{ACM-Reference-Format}
\bibliography{main}


\begin{thebibliography}{57}


\ifx \showCODEN    \undefined \def \showCODEN     #1{\unskip}     \fi
\ifx \showDOI      \undefined \def \showDOI       #1{#1}\fi
\ifx \showISBNx    \undefined \def \showISBNx     #1{\unskip}     \fi
\ifx \showISBNxiii \undefined \def \showISBNxiii  #1{\unskip}     \fi
\ifx \showISSN     \undefined \def \showISSN      #1{\unskip}     \fi
\ifx \showLCCN     \undefined \def \showLCCN      #1{\unskip}     \fi
\ifx \shownote     \undefined \def \shownote      #1{#1}          \fi
\ifx \showarticletitle \undefined \def \showarticletitle #1{#1}   \fi
\ifx \showURL      \undefined \def \showURL       {\relax}        \fi
\providecommand\bibfield[2]{#2}
\providecommand\bibinfo[2]{#2}
\providecommand\natexlab[1]{#1}
\providecommand\showeprint[2][]{arXiv:#2}

\bibitem[\protect\citeauthoryear{??}{WaP}{[n. d.]}]%
        {WaPost}
 \bibinfo{year}{[n. d.]}\natexlab{}.
\newblock \bibinfo{title}{About the Fact Checker}.
\newblock
  \bibinfo{howpublished}{\url{https://www.washingtonpost.com/politics/2019/01/07/about-fact-checker}}.
\newblock


\bibitem[\protect\citeauthoryear{??}{AMT}{[n. d.]}]%
        {AMTurk}
 \bibinfo{year}{[n. d.]}\natexlab{}.
\newblock \bibinfo{title}{Amazon Mechanical Turk}.
\newblock \bibinfo{howpublished}{\url{https://www.mturk.com}}.
\newblock


\bibitem[\protect\citeauthoryear{??}{BWd}{[n. d.]}]%
        {BWdata}
 \bibinfo{year}{[n. d.]}\natexlab{}.
\newblock \bibinfo{title}{BirdWatch Data}.
\newblock
  \bibinfo{howpublished}{\url{https://twitter.github.io/birdwatch/contributing/download-data/}}.
\newblock


\bibitem[\protect\citeauthoryear{??}{CRP}{[n. d.]}]%
        {CRProject}
 \bibinfo{year}{[n. d.]}\natexlab{}.
\newblock \bibinfo{title}{ClaimReview Project}.
\newblock
  \bibinfo{howpublished}{\url{https://www.claimreviewproject.com/the-facts-about-claimreview}}.
\newblock


\bibitem[\protect\citeauthoryear{??}{CRS}{[n. d.]}]%
        {CRSchema}
 \bibinfo{year}{[n. d.]}\natexlab{}.
\newblock \bibinfo{title}{ClaimReview Schema}.
\newblock \bibinfo{howpublished}{\url{https://schema.org/ClaimReview}}.
\newblock


\bibitem[\protect\citeauthoryear{??}{eva}{[n. d.]}]%
        {evaluate}
 \bibinfo{year}{[n. d.]}\natexlab{}.
\newblock \bibinfo{title}{Evaluating Evidence and Information Sources}.
\newblock
  \bibinfo{howpublished}{\url{https://kit.exposingtheinvisible.org/en/how/evaluate-evidence.html}}.
\newblock


\bibitem[\protect\citeauthoryear{??}{fff}{[n. d.]}]%
        {fffaq}
 \bibinfo{year}{[n. d.]}\natexlab{}.
\newblock \bibinfo{title}{Full Fact Frequently asked questions}.
\newblock
  \bibinfo{howpublished}{\url{https://fullfact.org/about/frequently-asked-questions/}}.
\newblock


\bibitem[\protect\citeauthoryear{??}{BWP}{[n. d.]}]%
        {BWPublish}
 \bibinfo{year}{[n. d.]}\natexlab{}.
\newblock \bibinfo{title}{Introducing Birdwatch, a community-based approach to
  misinformation}.
\newblock
  \bibinfo{howpublished}{\url{https://blog.twitter.com/en_us/topics/product/2021/introducing-birdwatch-a-community-based-approach-to-misinformation}}.
\newblock


\bibitem[\protect\citeauthoryear{??}{Tru}{[n. d.]}]%
        {TruthOMeter}
 \bibinfo{year}{[n. d.]}\natexlab{}.
\newblock \bibinfo{title}{The Principles of the Truth-O-Meter}.
\newblock
  \bibinfo{howpublished}{\url{ttps://www.politifact.com/article/2018/feb/12/principles-truth-o-meter-politifacts-methodology-i}}.
\newblock


\bibitem[\protect\citeauthoryear{??}{Fac}{[n. d.]}]%
        {FacebookProgram}
 \bibinfo{year}{[n. d.]}\natexlab{}.
\newblock \bibinfo{title}{Third party fact-checking}.
\newblock
  \bibinfo{howpublished}{\url{https://www.facebook.com/journalismproject/programs/third-party-fact-checking/how-it-works}}.
\newblock


\bibitem[\protect\citeauthoryear{Adler and Boscaini-Gilroy}{Adler and
  Boscaini-Gilroy}{2019}]%
        {adler2019}
\bibfield{author}{\bibinfo{person}{Ben Adler} {and} \bibinfo{person}{Giacomo
  Boscaini-Gilroy}.} \bibinfo{year}{2019}\natexlab{}.
\newblock \showarticletitle{Real-time {Claim} {Detection} from {News}
  {Articles} and {Retrieval} of {Semantically}-{Similar} {Factchecks}}. In
  \bibinfo{booktitle}{\emph{NewsIR’19 Workshop at SIGIR}}.
\newblock


\bibitem[\protect\citeauthoryear{Ahmadi, Sand, and Papotti}{Ahmadi
  et~al\mbox{.}}{2022}]%
        {AhmadICDE}
\bibfield{author}{\bibinfo{person}{Naser Ahmadi}, \bibinfo{person}{Hansjorg
  Sand}, {and} \bibinfo{person}{Paolo Papotti}.}
  \bibinfo{year}{2022}\natexlab{}.
\newblock \showarticletitle{Unsupervised Matching of Data and Text}. In
  \bibinfo{booktitle}{\emph{{ICDE}}}. \bibinfo{publisher}{{IEEE}},
  \bibinfo{pages}{1058--1070}.
\newblock
\urldef\tempurl%
\url{https://doi.org/10.1109/ICDE53745.2022.00084}
\showDOI{\tempurl}


\bibitem[\protect\citeauthoryear{Allen, Martel, and Rand}{Allen
  et~al\mbox{.}}{2021}]%
        {allen_martel_rand_2021}
\bibfield{author}{\bibinfo{person}{Jennifer N~L Allen},
  \bibinfo{person}{Cameron Martel}, {and} \bibinfo{person}{David~G Rand}.}
  \bibinfo{year}{2021}\natexlab{}.
\newblock \bibinfo{title}{Birds of a feather don’t fact-check each other:
  Partisanship and the evaluation of news in Twitter’s Birdwatch crowdsourced
  fact-checking program}.
\newblock
\newblock
\urldef\tempurl%
\url{https://doi.org/10.31234/osf.io/57e3q}
\showDOI{\tempurl}


\bibitem[\protect\citeauthoryear{Alshomary, D\"{u}sterhus, and
  Wachsmuth}{Alshomary et~al\mbox{.}}{2020}]%
        {Alshomary:20}
\bibfield{author}{\bibinfo{person}{Milad Alshomary}, \bibinfo{person}{Nick
  D\"{u}sterhus}, {and} \bibinfo{person}{Henning Wachsmuth}.}
  \bibinfo{year}{2020}\natexlab{}.
\newblock \showarticletitle{Extractive Snippet Generation for Arguments}. In
  \bibinfo{booktitle}{\emph{SIGIR}}. \bibinfo{pages}{1969–1972}.
\newblock


\bibitem[\protect\citeauthoryear{Arnold}{Arnold}{2020}]%
        {fullfact:coof}
\bibfield{author}{\bibinfo{person}{Phoebe Arnold}.}
  \bibinfo{year}{2020}\natexlab{}.
\newblock \bibinfo{booktitle}{\emph{The challenges of online fact checking}}.
\newblock \bibinfo{type}{{T}echnical {R}eport}. \bibinfo{institution}{Full
  Fact}.
\newblock


\bibitem[\protect\citeauthoryear{Atanasova, Nakov, Karadzhov, Mohtarami, and
  Martino}{Atanasova et~al\mbox{.}}{2019}]%
        {clef-checkthat-T1:2019}
\bibfield{author}{\bibinfo{person}{Pepa Atanasova}, \bibinfo{person}{Preslav
  Nakov}, \bibinfo{person}{Georgi Karadzhov}, \bibinfo{person}{Mitra
  Mohtarami}, {and} \bibinfo{person}{Giovanni Da~San Martino}.}
  \bibinfo{year}{2019}\natexlab{}.
\newblock \showarticletitle{Overview of the {CLEF-2019 CheckThat! Lab on
  Automatic Identification and Verification of Claims. Task 1:
  Check-Worthiness}}. In \bibinfo{booktitle}{\emph{Working Notes of CLEF}}
  \emph{(\bibinfo{series}{CLEF~'19})}.
\newblock


\bibitem[\protect\citeauthoryear{Barbera, Roitero, Demartini, Mizzaro, and
  Spina}{Barbera et~al\mbox{.}}{2020}]%
        {DBLP:conf/ecir/BarberaRDMS20}
\bibfield{author}{\bibinfo{person}{David~La Barbera}, \bibinfo{person}{Kevin
  Roitero}, \bibinfo{person}{Gianluca Demartini}, \bibinfo{person}{Stefano
  Mizzaro}, {and} \bibinfo{person}{Damiano Spina}.}
  \bibinfo{year}{2020}\natexlab{}.
\newblock \showarticletitle{Crowdsourcing Truthfulness: The Impact of Judgment
  Scale and Assessor Bias}. In \bibinfo{booktitle}{\emph{{ECIR}}}
  \emph{(\bibinfo{series}{Lecture Notes in Computer Science})},
  Vol.~\bibinfo{volume}{12036}. \bibinfo{publisher}{Springer},
  \bibinfo{pages}{207--214}.
\newblock
\urldef\tempurl%
\url{https://doi.org/10.1007/978-3-030-45442-5\_26}
\showDOI{\tempurl}


\bibitem[\protect\citeauthoryear{Bhuiyan, Zhang, Sehat, and Mitra}{Bhuiyan
  et~al\mbox{.}}{2020}]%
        {BhuiyanZSM20}
\bibfield{author}{\bibinfo{person}{Md~Momen Bhuiyan}, \bibinfo{person}{Amy~X.
  Zhang}, \bibinfo{person}{Connie~Moon Sehat}, {and} \bibinfo{person}{Tanushree
  Mitra}.} \bibinfo{year}{2020}\natexlab{}.
\newblock \showarticletitle{Investigating Differences in Crowdsourced News
  Credibility Assessment: Raters, Tasks, and Expert Criteria}.
\newblock \bibinfo{journal}{\emph{Proc. {ACM} Hum. Comput. Interact.}}
  \bibinfo{volume}{4}, \bibinfo{number}{{CSCW2}} (\bibinfo{year}{2020}),
  \bibinfo{pages}{93:1--93:26}.
\newblock
\urldef\tempurl%
\url{https://doi.org/10.1145/3415164}
\showDOI{\tempurl}


\bibitem[\protect\citeauthoryear{Botnevik, Sakariassen, and Setty}{Botnevik
  et~al\mbox{.}}{2020}]%
        {botnevik2020brenda}
\bibfield{author}{\bibinfo{person}{Bjarte Botnevik}, \bibinfo{person}{Eirik
  Sakariassen}, {and} \bibinfo{person}{Vinay Setty}.}
  \bibinfo{year}{2020}\natexlab{}.
\newblock \showarticletitle{{BRENDA}: Browser Extension for Fake News
  Detection}. In \bibinfo{booktitle}{\emph{SIGIR}}.
  \bibinfo{pages}{2117--2120}.
\newblock


\bibitem[\protect\citeauthoryear{Brand, Roitero, Soprano, and Demartini}{Brand
  et~al\mbox{.}}{2021}]%
        {BrandRSD21}
\bibfield{author}{\bibinfo{person}{Erik Brand}, \bibinfo{person}{Kevin
  Roitero}, \bibinfo{person}{Michael Soprano}, {and} \bibinfo{person}{Gianluca
  Demartini}.} \bibinfo{year}{2021}\natexlab{}.
\newblock \showarticletitle{{E-BART:} Jointly Predicting and Explaining
  Truthfulness}. In \bibinfo{booktitle}{\emph{{TTO}}}. \bibinfo{pages}{18--27}.
\newblock


\bibitem[\protect\citeauthoryear{Checco, Roitero, Maddalena, Mizzaro, and
  Demartini}{Checco et~al\mbox{.}}{2017}]%
        {checco2017let}
\bibfield{author}{\bibinfo{person}{Alessandro Checco}, \bibinfo{person}{Kevin
  Roitero}, \bibinfo{person}{Eddy Maddalena}, \bibinfo{person}{Stefano
  Mizzaro}, {and} \bibinfo{person}{Gianluca Demartini}.}
  \bibinfo{year}{2017}\natexlab{}.
\newblock \showarticletitle{Let's agree to disagree: Fixing agreement measures
  for crowdsourcing}. In \bibinfo{booktitle}{\emph{Fifth AAAI Conference on
  Human Computation and Crowdsourcing}}.
\newblock


\bibitem[\protect\citeauthoryear{Dori-Hacohen, Jensen, and Allan}{Dori-Hacohen
  et~al\mbox{.}}{2016}]%
        {10.1145/2911451.2914745}
\bibfield{author}{\bibinfo{person}{Shiri Dori-Hacohen}, \bibinfo{person}{David
  Jensen}, {and} \bibinfo{person}{James Allan}.}
  \bibinfo{year}{2016}\natexlab{}.
\newblock \showarticletitle{Controversy Detection in Wikipedia Using Collective
  Classification}. In \bibinfo{booktitle}{\emph{Proceedings of the 39th
  International ACM SIGIR Conference on Research and Development in Information
  Retrieval}} \emph{(\bibinfo{series}{SIGIR '16})}.
  \bibinfo{publisher}{Association for Computing Machinery},
  \bibinfo{address}{New York, NY, USA}, \bibinfo{pages}{797–800}.
\newblock
\showISBNx{9781450340694}
\urldef\tempurl%
\url{https://doi.org/10.1145/2911451.2914745}
\showDOI{\tempurl}


\bibitem[\protect\citeauthoryear{Dori-Hacohen, Sung, Chou, and
  Lustig-Gonzalez}{Dori-Hacohen et~al\mbox{.}}{2021}]%
        {10.1145/3404835.3464926}
\bibfield{author}{\bibinfo{person}{Shiri Dori-Hacohen}, \bibinfo{person}{Keen
  Sung}, \bibinfo{person}{Jengyu Chou}, {and} \bibinfo{person}{Julian
  Lustig-Gonzalez}.} \bibinfo{year}{2021}\natexlab{}.
\newblock \bibinfo{booktitle}{\emph{Restoring Healthy Online Discourse by
  Detecting and Reducing Controversy, Misinformation, and Toxicity Online}}.
\newblock \bibinfo{publisher}{Association for Computing Machinery},
  \bibinfo{address}{New York, NY, USA}, \bibinfo{pages}{2627–2628}.
\newblock
\showISBNx{9781450380379}
\urldef\tempurl%
\url{https://doi.org/10.1145/3404835.3464926}
\showURL{%
\tempurl}


\bibitem[\protect\citeauthoryear{Epstein, Pennycook, and Rand}{Epstein
  et~al\mbox{.}}{2020}]%
        {EpsteinPR20}
\bibfield{author}{\bibinfo{person}{Ziv Epstein}, \bibinfo{person}{Gordon
  Pennycook}, {and} \bibinfo{person}{David~G. Rand}.}
  \bibinfo{year}{2020}\natexlab{}.
\newblock \showarticletitle{Will the Crowd Game the Algorithm?: Using Layperson
  Judgments to Combat Misinformation on Social Media by Downranking Distrusted
  Sources}. In \bibinfo{booktitle}{\emph{{CHI}}}. \bibinfo{publisher}{{ACM}},
  \bibinfo{pages}{1--11}.
\newblock
\urldef\tempurl%
\url{https://doi.org/10.1145/3313831.3376232}
\showDOI{\tempurl}


\bibitem[\protect\citeauthoryear{Fan, Piktus, Petroni, Wenzek, Saeidi, Vlachos,
  Bordes, and Riedel}{Fan et~al\mbox{.}}{2020}]%
        {fan2020generating}
\bibfield{author}{\bibinfo{person}{Angela Fan}, \bibinfo{person}{Aleksandra
  Piktus}, \bibinfo{person}{Fabio Petroni}, \bibinfo{person}{Guillaume Wenzek},
  \bibinfo{person}{Marzieh Saeidi}, \bibinfo{person}{Andreas Vlachos},
  \bibinfo{person}{Antoine Bordes}, {and} \bibinfo{person}{Sebastian Riedel}.}
  \bibinfo{year}{2020}\natexlab{}.
\newblock \showarticletitle{Generating Fact Checking Briefs}. In
  \bibinfo{booktitle}{\emph{Proceedings of the 2020 Conference on Empirical
  Methods in Natural Language Processing (EMNLP)}}.
  \bibinfo{pages}{7147--7161}.
\newblock


\bibitem[\protect\citeauthoryear{Fleiss}{Fleiss}{1971}]%
        {Fleiss1971MeasuringNS}
\bibfield{author}{\bibinfo{person}{Joseph~L. Fleiss}.}
  \bibinfo{year}{1971}\natexlab{}.
\newblock \showarticletitle{Measuring nominal scale agreement among many
  raters.}
\newblock \bibinfo{journal}{\emph{Psychological Bulletin}}
  \bibinfo{volume}{76} (\bibinfo{year}{1971}), \bibinfo{pages}{378--382}.
\newblock


\bibitem[\protect\citeauthoryear{Grootendorst}{Grootendorst}{2020}]%
        {grootendorst2020bertopic}
\bibfield{author}{\bibinfo{person}{Maarten Grootendorst}.}
  \bibinfo{year}{2020}\natexlab{}.
\newblock \bibinfo{title}{BERTopic: Leveraging BERT and c-TF-IDF to create
  easily interpretable topics.}
\newblock
\newblock
\urldef\tempurl%
\url{https://doi.org/10.5281/zenodo.4381785}
\showDOI{\tempurl}


\bibitem[\protect\citeauthoryear{Hansen, Hansen, Alstrup, Grue~Simonsen, and
  Lioma}{Hansen et~al\mbox{.}}{2019}]%
        {10.1145/3308560.3316736}
\bibfield{author}{\bibinfo{person}{Casper Hansen}, \bibinfo{person}{Christian
  Hansen}, \bibinfo{person}{Stephen Alstrup}, \bibinfo{person}{Jakob
  Grue~Simonsen}, {and} \bibinfo{person}{Christina Lioma}.}
  \bibinfo{year}{2019}\natexlab{}.
\newblock \showarticletitle{Neural Check-Worthiness Ranking with Weak
  Supervision: Finding Sentences for Fact-Checking}. In
  \bibinfo{booktitle}{\emph{Companion Proceedings of The 2019 World Wide Web
  Conference}} \emph{(\bibinfo{series}{WWW '19})}.
  \bibinfo{publisher}{Association for Computing Machinery},
  \bibinfo{address}{New York, NY, USA}, \bibinfo{pages}{994–1000}.
\newblock
\showISBNx{9781450366755}
\urldef\tempurl%
\url{https://doi.org/10.1145/3308560.3316736}
\showDOI{\tempurl}


\bibitem[\protect\citeauthoryear{Hassan, Adair, Hamilton, Li, Tremayne, Yang,
  and Yu}{Hassan et~al\mbox{.}}{2015}]%
        {hassan2015quest}
\bibfield{author}{\bibinfo{person}{Naeemul Hassan}, \bibinfo{person}{Bill
  Adair}, \bibinfo{person}{James~T Hamilton}, \bibinfo{person}{Chengkai Li},
  \bibinfo{person}{Mark Tremayne}, \bibinfo{person}{Jun Yang}, {and}
  \bibinfo{person}{Cong Yu}.} \bibinfo{year}{2015}\natexlab{}.
\newblock \showarticletitle{The quest to automate fact-checking}. In
  \bibinfo{booktitle}{\emph{Proceedings of the 2015 computation+ journalism
  symposium}}.
\newblock


\bibitem[\protect\citeauthoryear{Hassan, Arslan, Li, and Tremayne}{Hassan
  et~al\mbox{.}}{2017a}]%
        {HassanALT17}
\bibfield{author}{\bibinfo{person}{Naeemul Hassan}, \bibinfo{person}{Fatma
  Arslan}, \bibinfo{person}{Chengkai Li}, {and} \bibinfo{person}{Mark
  Tremayne}.} \bibinfo{year}{2017}\natexlab{a}.
\newblock \showarticletitle{Toward Automated Fact-Checking: Detecting
  Check-worthy Factual Claims by ClaimBuster}. In
  \bibinfo{booktitle}{\emph{{KDD}}}.
\newblock


\bibitem[\protect\citeauthoryear{Hassan, Yousuf, Mahfuzul~Haque,
  A.~Suarez~Rivas, and Khadimul~Islam}{Hassan et~al\mbox{.}}{2019}]%
        {10.1145/3308560.3316734}
\bibfield{author}{\bibinfo{person}{Naeemul Hassan}, \bibinfo{person}{Mohammad
  Yousuf}, \bibinfo{person}{Md Mahfuzul~Haque}, \bibinfo{person}{Javier
  A.~Suarez~Rivas}, {and} \bibinfo{person}{Md Khadimul~Islam}.}
  \bibinfo{year}{2019}\natexlab{}.
\newblock \showarticletitle{Examining the Roles of Automation, Crowds and
  Professionals Towards Sustainable Fact-Checking}. In
  \bibinfo{booktitle}{\emph{Companion Proceedings of The 2019 World Wide Web
  Conference}} \emph{(\bibinfo{series}{WWW '19})}.
  \bibinfo{publisher}{Association for Computing Machinery},
  \bibinfo{address}{New York, NY, USA}, \bibinfo{pages}{1001–1006}.
\newblock
\showISBNx{9781450366755}
\urldef\tempurl%
\url{https://doi.org/10.1145/3308560.3316734}
\showDOI{\tempurl}


\bibitem[\protect\citeauthoryear{Hassan, Zhang, Arslan, Caraballo, Jimenez,
  Gawsane, Hasan, Joseph, Kulkarni, Nayak, Sable, Li, and Tremayne}{Hassan
  et~al\mbox{.}}{2017b}]%
        {HassanZACJGHJKN17}
\bibfield{author}{\bibinfo{person}{Naeemul Hassan}, \bibinfo{person}{Gensheng
  Zhang}, \bibinfo{person}{Fatma Arslan}, \bibinfo{person}{Josue Caraballo},
  \bibinfo{person}{Damian Jimenez}, \bibinfo{person}{Siddhant Gawsane},
  \bibinfo{person}{Shohedul Hasan}, \bibinfo{person}{Minumol Joseph},
  \bibinfo{person}{Aaditya Kulkarni}, \bibinfo{person}{Anil~Kumar Nayak},
  \bibinfo{person}{Vikas Sable}, \bibinfo{person}{Chengkai Li}, {and}
  \bibinfo{person}{Mark Tremayne}.} \bibinfo{year}{2017}\natexlab{b}.
\newblock \showarticletitle{ClaimBuster: The First-ever End-to-end
  Fact-checking System}.
\newblock \bibinfo{journal}{\emph{Proc. {VLDB} Endow.}} \bibinfo{volume}{10},
  \bibinfo{number}{12} (\bibinfo{year}{2017}), \bibinfo{pages}{1945--1948}.
\newblock
\urldef\tempurl%
\url{https://doi.org/10.14778/3137765.3137815}
\showDOI{\tempurl}


\bibitem[\protect\citeauthoryear{Huynh, Nguyen, Goh, Kim, and Hong}{Huynh
  et~al\mbox{.}}{2021}]%
        {huynh2021argh}
\bibfield{author}{\bibinfo{person}{Larry Huynh}, \bibinfo{person}{Thai Nguyen},
  \bibinfo{person}{Joshua Goh}, \bibinfo{person}{Hyoungshick Kim}, {and}
  \bibinfo{person}{Jin~B Hong}.} \bibinfo{year}{2021}\natexlab{}.
\newblock \showarticletitle{ARGH! Automated Rumor Generation Hub}. In
  \bibinfo{booktitle}{\emph{Proceedings of the 30th ACM International
  Conference on Information \& Knowledge Management}}.
  \bibinfo{pages}{3847--3856}.
\newblock


\bibitem[\protect\citeauthoryear{Karagiannis, Saeed, Papotti, and
  Trummer}{Karagiannis et~al\mbox{.}}{2020}]%
        {Karagiannis0PT20}
\bibfield{author}{\bibinfo{person}{Georgios Karagiannis},
  \bibinfo{person}{Mohammed Saeed}, \bibinfo{person}{Paolo Papotti}, {and}
  \bibinfo{person}{Immanuel Trummer}.} \bibinfo{year}{2020}\natexlab{}.
\newblock \showarticletitle{Scrutinizer: {A} Mixed-Initiative Approach to
  Large-Scale, Data-Driven Claim Verification}.
\newblock \bibinfo{journal}{\emph{Proc. {VLDB} Endow.}} \bibinfo{volume}{13},
  \bibinfo{number}{11} (\bibinfo{year}{2020}), \bibinfo{pages}{2508--2521}.
\newblock


\bibitem[\protect\citeauthoryear{Kazai}{Kazai}{2011}]%
        {Kazai2011}
\bibfield{author}{\bibinfo{person}{Gabriella Kazai}.}
  \bibinfo{year}{2011}\natexlab{}.
\newblock \showarticletitle{In Search of Quality in Crowdsourcing for Search
  Engine Evaluation}. In \bibinfo{booktitle}{\emph{Proceedings of the 33rd
  European Conference on Advances in Information Retrieval - Volume 6611}}
  \emph{(\bibinfo{series}{ECIR 2011})}. \bibinfo{publisher}{Springer-Verlag},
  \bibinfo{address}{Berlin, Heidelberg}, \bibinfo{pages}{165–176}.
\newblock
\showISBNx{9783642201608}
\urldef\tempurl%
\url{https://doi.org/10.1007/978-3-642-20161-5_17}
\showDOI{\tempurl}


\bibitem[\protect\citeauthoryear{Krippendorff}{Krippendorff}{2011}]%
        {Krippendorff2011ComputingKA}
\bibfield{author}{\bibinfo{person}{Klaus Krippendorff}.}
  \bibinfo{year}{2011}\natexlab{}.
\newblock \bibinfo{booktitle}{\emph{Computing Krippendorff's
  Alpha-Reliability}}.
\newblock \bibinfo{type}{{T}echnical {R}eport}.
\newblock


\bibitem[\protect\citeauthoryear{Liu, Bias, Lease, and Kuipers}{Liu
  et~al\mbox{.}}{2012}]%
        {Du2012}
\bibfield{author}{\bibinfo{person}{Di Liu}, \bibinfo{person}{Randolph~G. Bias},
  \bibinfo{person}{Matthew Lease}, {and} \bibinfo{person}{Rebecca Kuipers}.}
  \bibinfo{year}{2012}\natexlab{}.
\newblock \showarticletitle{Crowdsourcing for usability testing}.
\newblock \bibinfo{journal}{\emph{Proceedings of the American Society for
  Information Science and Technology}} \bibinfo{volume}{49},
  \bibinfo{number}{1} (\bibinfo{year}{2012}), \bibinfo{pages}{1--10}.
\newblock
\urldef\tempurl%
\url{https://doi.org/10.1002/meet.14504901100}
\showDOI{\tempurl}
\showeprint{https://asistdl.onlinelibrary.wiley.com/doi/pdf/10.1002/meet.14504901100}


\bibitem[\protect\citeauthoryear{Liu and fang Brook~Wu}{Liu and fang
  Brook~Wu}{2018}]%
        {Liu2018EarlyDO}
\bibfield{author}{\bibinfo{person}{Yang~P. Liu} {and} \bibinfo{person}{Yi fang
  Brook~Wu}.} \bibinfo{year}{2018}\natexlab{}.
\newblock \showarticletitle{Early Detection of Fake News on Social Media
  Through Propagation Path Classification with Recurrent and Convolutional
  Networks}. In \bibinfo{booktitle}{\emph{AAAI}}.
\newblock


\bibitem[\protect\citeauthoryear{Mensio and Alani}{Mensio and Alani}{2019}]%
        {MensioA19}
\bibfield{author}{\bibinfo{person}{Martino Mensio} {and}
  \bibinfo{person}{Harith Alani}.} \bibinfo{year}{2019}\natexlab{}.
\newblock \showarticletitle{MisinfoMe: Who is Interacting with
  Misinformation?}. In \bibinfo{booktitle}{\emph{{ISWC}}}
  \emph{(\bibinfo{series}{{CEUR} Workshop Proceedings})},
  Vol.~\bibinfo{volume}{2456}. \bibinfo{publisher}{CEUR-WS.org},
  \bibinfo{pages}{217--220}.
\newblock


\bibitem[\protect\citeauthoryear{Micallef, He, Kumar, Ahamad, and
  Memon}{Micallef et~al\mbox{.}}{2020}]%
        {9377956}
\bibfield{author}{\bibinfo{person}{Nicholas Micallef}, \bibinfo{person}{Bing
  He}, \bibinfo{person}{Srijan Kumar}, \bibinfo{person}{Mustaque Ahamad}, {and}
  \bibinfo{person}{Nasir Memon}.} \bibinfo{year}{2020}\natexlab{}.
\newblock \showarticletitle{The Role of the Crowd in Countering Misinformation:
  A Case Study of the COVID-19 Infodemic}. In \bibinfo{booktitle}{\emph{2020
  IEEE International Conference on Big Data (Big Data)}}.
  \bibinfo{pages}{748--757}.
\newblock
\urldef\tempurl%
\url{https://doi.org/10.1109/BigData50022.2020.9377956}
\showDOI{\tempurl}


\bibitem[\protect\citeauthoryear{Nakov, Corney, Hasanain, Alam, Elsayed,
  Barr{\'o}n-Cede{\~n}o, Papotti, Shaar, and Martino}{Nakov
  et~al\mbox{.}}{2021}]%
        {nakov2021automated}
\bibfield{author}{\bibinfo{person}{Preslav Nakov}, \bibinfo{person}{David
  Corney}, \bibinfo{person}{Maram Hasanain}, \bibinfo{person}{Firoj Alam},
  \bibinfo{person}{Tamer Elsayed}, \bibinfo{person}{Alberto
  Barr{\'o}n-Cede{\~n}o}, \bibinfo{person}{Paolo Papotti},
  \bibinfo{person}{Shaden Shaar}, {and} \bibinfo{person}{Giovanni Da~San
  Martino}.} \bibinfo{year}{2021}\natexlab{}.
\newblock \showarticletitle{Automated fact-checking for assisting human
  fact-checkers}.
\newblock \bibinfo{journal}{\emph{IJCAI}} (\bibinfo{year}{2021}).
\newblock


\bibitem[\protect\citeauthoryear{Nevo and Horne}{Nevo and Horne}{2022}]%
        {Nevo2022}
\bibfield{author}{\bibinfo{person}{Dorit Nevo} {and}
  \bibinfo{person}{Benjamin~D. Horne}.} \bibinfo{year}{2022}\natexlab{}.
\newblock \showarticletitle{How Topic Novelty Impacts the Effectiveness of News
  Veracity Interventions}.
\newblock \bibinfo{journal}{\emph{Commun. ACM}} \bibinfo{volume}{65},
  \bibinfo{number}{2} (\bibinfo{date}{jan} \bibinfo{year}{2022}),
  \bibinfo{pages}{68–75}.
\newblock
\showISSN{0001-0782}
\urldef\tempurl%
\url{https://doi.org/10.1145/3460350}
\showDOI{\tempurl}


\bibitem[\protect\citeauthoryear{Nguyen, Kharosekar, Krishnan, Krishnan, Tate,
  Wallace, and Lease}{Nguyen et~al\mbox{.}}{2018}]%
        {10.1145/3242587.3242666}
\bibfield{author}{\bibinfo{person}{An~T. Nguyen}, \bibinfo{person}{Aditya
  Kharosekar}, \bibinfo{person}{Saumyaa Krishnan}, \bibinfo{person}{Siddhesh
  Krishnan}, \bibinfo{person}{Elizabeth Tate}, \bibinfo{person}{Byron~C.
  Wallace}, {and} \bibinfo{person}{Matthew Lease}.}
  \bibinfo{year}{2018}\natexlab{}.
\newblock \showarticletitle{Believe It or Not: Designing a Human-AI Partnership
  for Mixed-Initiative Fact-Checking}. In \bibinfo{booktitle}{\emph{Proceedings
  of the 31st Annual ACM Symposium on User Interface Software and Technology}}
  \emph{(\bibinfo{series}{UIST '18})}. \bibinfo{publisher}{Association for
  Computing Machinery}, \bibinfo{address}{New York, NY, USA},
  \bibinfo{pages}{189–199}.
\newblock
\showISBNx{9781450359481}
\urldef\tempurl%
\url{https://doi.org/10.1145/3242587.3242666}
\showDOI{\tempurl}


\bibitem[\protect\citeauthoryear{Pinto, de~Lima, Barbosa, and de~Souza}{Pinto
  et~al\mbox{.}}{2019}]%
        {8791903}
\bibfield{author}{\bibinfo{person}{Marcos~Rodrigues Pinto},
  \bibinfo{person}{Yuri~Oliveira de Lima}, \bibinfo{person}{Carlos~Eduardo
  Barbosa}, {and} \bibinfo{person}{Jano~Moreira de Souza}.}
  \bibinfo{year}{2019}\natexlab{}.
\newblock \showarticletitle{Towards Fact-Checking through Crowdsourcing}. In
  \bibinfo{booktitle}{\emph{2019 IEEE 23rd International Conference on Computer
  Supported Cooperative Work in Design (CSCWD)}}. \bibinfo{pages}{494--499}.
\newblock
\urldef\tempurl%
\url{https://doi.org/10.1109/CSCWD.2019.8791903}
\showDOI{\tempurl}


\bibitem[\protect\citeauthoryear{Pradeep, Ma, Nogueira, and Lin}{Pradeep
  et~al\mbox{.}}{2021}]%
        {10.1145/3404835.3463120}
\bibfield{author}{\bibinfo{person}{Ronak Pradeep}, \bibinfo{person}{Xueguang
  Ma}, \bibinfo{person}{Rodrigo Nogueira}, {and} \bibinfo{person}{Jimmy Lin}.}
  \bibinfo{year}{2021}\natexlab{}.
\newblock \bibinfo{booktitle}{\emph{Vera: Prediction Techniques for Reducing
  Harmful Misinformation in Consumer Health Search}}.
\newblock \bibinfo{publisher}{Association for Computing Machinery},
  \bibinfo{address}{New York, NY, USA}, \bibinfo{pages}{2066–2070}.
\newblock
\showISBNx{9781450380379}
\urldef\tempurl%
\url{https://doi.org/10.1145/3404835.3463120}
\showURL{%
\tempurl}


\bibitem[\protect\citeauthoryear{Pr{\"{o}}llochs}{Pr{\"{o}}llochs}{2021}]%
        {Prollochs2021CommunityBasedFO}
\bibfield{author}{\bibinfo{person}{Nicolas Pr{\"{o}}llochs}.}
  \bibinfo{year}{2021}\natexlab{}.
\newblock \showarticletitle{Community-Based Fact-Checking on Twitter's
  Birdwatch Platform}.
\newblock \bibinfo{journal}{\emph{CoRR}}  \bibinfo{volume}{abs/2104.07175}
  (\bibinfo{year}{2021}).
\newblock
\showeprint[arXiv]{2104.07175}
\urldef\tempurl%
\url{https://arxiv.org/abs/2104.07175}
\showURL{%
\tempurl}


\bibitem[\protect\citeauthoryear{Reimers and Gurevych}{Reimers and
  Gurevych}{2019}]%
        {reimers-gurevych-2019-sentence}
\bibfield{author}{\bibinfo{person}{Nils Reimers} {and} \bibinfo{person}{Iryna
  Gurevych}.} \bibinfo{year}{2019}\natexlab{}.
\newblock \showarticletitle{Sentence-{BERT}: Sentence Embeddings using
  {S}iamese {BERT}-Networks}. In \bibinfo{booktitle}{\emph{Proceedings of the
  2019 Conference on Empirical Methods in Natural Language Processing and the
  9th International Joint Conference on Natural Language Processing
  (EMNLP-IJCNLP)}}. \bibinfo{publisher}{Association for Computational
  Linguistics}, \bibinfo{address}{Hong Kong, China},
  \bibinfo{pages}{3982--3992}.
\newblock
\urldef\tempurl%
\url{https://doi.org/10.18653/v1/D19-1410}
\showDOI{\tempurl}


\bibitem[\protect\citeauthoryear{Roitero, Soprano, Fan, Spina, Mizzaro, and
  Demartini}{Roitero et~al\mbox{.}}{2020a}]%
        {roitero2020can}
\bibfield{author}{\bibinfo{person}{Kevin Roitero}, \bibinfo{person}{Michael
  Soprano}, \bibinfo{person}{Shaoyang Fan}, \bibinfo{person}{Damiano Spina},
  \bibinfo{person}{Stefano Mizzaro}, {and} \bibinfo{person}{Gianluca
  Demartini}.} \bibinfo{year}{2020}\natexlab{a}.
\newblock \showarticletitle{Can The Crowd Identify Misinformation Objectively?
  The Effects of Judgment Scale and Assessor's Background}. In
  \bibinfo{booktitle}{\emph{Proceedings of the 43rd International ACM SIGIR
  Conference on Research and Development in Information Retrieval}}.
  \bibinfo{pages}{439--448}.
\newblock


\bibitem[\protect\citeauthoryear{Roitero, Soprano, Portelli, Spina, Della~Mea,
  Serra, Mizzaro, and Demartini}{Roitero et~al\mbox{.}}{2020b}]%
        {10.1145/3340531.3412048}
\bibfield{author}{\bibinfo{person}{Kevin Roitero}, \bibinfo{person}{Michael
  Soprano}, \bibinfo{person}{Beatrice Portelli}, \bibinfo{person}{Damiano
  Spina}, \bibinfo{person}{Vincenzo Della~Mea}, \bibinfo{person}{Giuseppe
  Serra}, \bibinfo{person}{Stefano Mizzaro}, {and} \bibinfo{person}{Gianluca
  Demartini}.} \bibinfo{year}{2020}\natexlab{b}.
\newblock \showarticletitle{The COVID-19 Infodemic: Can the Crowd Judge Recent
  Misinformation Objectively?}. In \bibinfo{booktitle}{\emph{{CIKM}}}
  \emph{(\bibinfo{series}{CIKM '20})}. \bibinfo{publisher}{Association for
  Computing Machinery}, \bibinfo{address}{New York, NY, USA},
  \bibinfo{pages}{1305–1314}.
\newblock
\showISBNx{9781450368599}
\urldef\tempurl%
\url{https://doi.org/10.1145/3340531.3412048}
\showDOI{\tempurl}


\bibitem[\protect\citeauthoryear{Shaar, Babulkov, Martino, and Nakov}{Shaar
  et~al\mbox{.}}{2020}]%
        {ShaarBMN20}
\bibfield{author}{\bibinfo{person}{Shaden Shaar}, \bibinfo{person}{Nikolay
  Babulkov}, \bibinfo{person}{Giovanni Da~San Martino}, {and}
  \bibinfo{person}{Preslav Nakov}.} \bibinfo{year}{2020}\natexlab{}.
\newblock \showarticletitle{That is a Known Lie: Detecting Previously
  Fact-Checked Claims}. In \bibinfo{booktitle}{\emph{{ACL}}}.
  \bibinfo{pages}{3607--3618}.
\newblock


\bibitem[\protect\citeauthoryear{Shin, Jian, Driscoll, and Bar}{Shin
  et~al\mbox{.}}{2017}]%
        {shin2017}
\bibfield{author}{\bibinfo{person}{Jieun Shin}, \bibinfo{person}{Lian Jian},
  \bibinfo{person}{Kevin Driscoll}, {and} \bibinfo{person}{François Bar}.}
  \bibinfo{year}{2017}\natexlab{}.
\newblock \showarticletitle{Political rumoring on Twitter during the 2012 US
  presidential election: Rumor diffusion and correction}.
\newblock \bibinfo{journal}{\emph{New Media \& Society}} \bibinfo{volume}{19},
  \bibinfo{number}{8} (\bibinfo{year}{2017}), \bibinfo{pages}{1214--1235}.
\newblock
\urldef\tempurl%
\url{https://doi.org/10.1177/1461444816634054}
\showDOI{\tempurl}
\showeprint{https://doi.org/10.1177/1461444816634054}


\bibitem[\protect\citeauthoryear{Starbird}{Starbird}{2019}]%
        {starbird2019disinformation}
\bibfield{author}{\bibinfo{person}{Kate Starbird}.}
  \bibinfo{year}{2019}\natexlab{}.
\newblock \showarticletitle{Disinformation's spread: bots, trolls and all of
  us}.
\newblock \bibinfo{journal}{\emph{Nature}} \bibinfo{volume}{571},
  \bibinfo{number}{7766} (\bibinfo{year}{2019}), \bibinfo{pages}{449--450}.
\newblock


\bibitem[\protect\citeauthoryear{Su, Macdonald, and Ounis}{Su
  et~al\mbox{.}}{2019}]%
        {10.1145/3331184.3331305}
\bibfield{author}{\bibinfo{person}{Ting Su}, \bibinfo{person}{Craig Macdonald},
  {and} \bibinfo{person}{Iadh Ounis}.} \bibinfo{year}{2019}\natexlab{}.
\newblock \showarticletitle{Ensembles of Recurrent Networks for Classifying the
  Relationship of Fake News Titles}. In \bibinfo{booktitle}{\emph{Proceedings
  of the 42nd International ACM SIGIR Conference on Research and Development in
  Information Retrieval}} \emph{(\bibinfo{series}{SIGIR'19})}.
  \bibinfo{publisher}{Association for Computing Machinery},
  \bibinfo{address}{New York, NY, USA}, \bibinfo{pages}{893–896}.
\newblock
\showISBNx{9781450361729}
\urldef\tempurl%
\url{https://doi.org/10.1145/3331184.3331305}
\showDOI{\tempurl}


\bibitem[\protect\citeauthoryear{Thorne and Vlachos}{Thorne and
  Vlachos}{2018}]%
        {thorne2018automated}
\bibfield{author}{\bibinfo{person}{James Thorne} {and} \bibinfo{person}{Andreas
  Vlachos}.} \bibinfo{year}{2018}\natexlab{}.
\newblock \showarticletitle{Automated Fact Checking: Task Formulations, Methods
  and Future Directions}. In \bibinfo{booktitle}{\emph{Proceedings of the 27th
  International Conference on Computational Linguistics}}.
  \bibinfo{pages}{3346--3359}.
\newblock


\bibitem[\protect\citeauthoryear{Thorne, Vlachos, Cocarascu,
  Christodoulopoulos, and Mittal}{Thorne et~al\mbox{.}}{2018}]%
        {thorne-etal-2018-fact}
\bibfield{author}{\bibinfo{person}{James Thorne}, \bibinfo{person}{Andreas
  Vlachos}, \bibinfo{person}{Oana Cocarascu}, \bibinfo{person}{Christos
  Christodoulopoulos}, {and} \bibinfo{person}{Arpit Mittal}.}
  \bibinfo{year}{2018}\natexlab{}.
\newblock \showarticletitle{The Fact Extraction and {VER}ification ({FEVER})
  Shared Task}. In \bibinfo{booktitle}{\emph{{FEVER}}}. \bibinfo{pages}{1--9}.
\newblock
\urldef\tempurl%
\url{https://doi.org/10.18653/v1/W18-5501}
\showDOI{\tempurl}


\bibitem[\protect\citeauthoryear{Vo and Lee}{Vo and Lee}{2018}]%
        {vo2018rise}
\bibfield{author}{\bibinfo{person}{Nguyen Vo} {and} \bibinfo{person}{Kyumin
  Lee}.} \bibinfo{year}{2018}\natexlab{}.
\newblock \showarticletitle{The rise of guardians: Fact-checking {URL}
  recommendation to combat fake news}. In \bibinfo{booktitle}{\emph{SIGIR}}.
  \bibinfo{pages}{275--284}.
\newblock


\bibitem[\protect\citeauthoryear{Wei, Xu, and Mao}{Wei et~al\mbox{.}}{2019}]%
        {wei-etal-2019-modeling}
\bibfield{author}{\bibinfo{person}{Penghui Wei}, \bibinfo{person}{Nan Xu},
  {and} \bibinfo{person}{Wenji Mao}.} \bibinfo{year}{2019}\natexlab{}.
\newblock \showarticletitle{Modeling Conversation Structure and Temporal
  Dynamics for Jointly Predicting Rumor Stance and Veracity}. In
  \bibinfo{booktitle}{\emph{{EMNLP-IJCNLP}}}. \bibinfo{publisher}{ACL},
  \bibinfo{pages}{4787--4798}.
\newblock
\urldef\tempurl%
\url{https://doi.org/10.18653/v1/D19-1485}
\showDOI{\tempurl}


\end{thebibliography}

\balance

\end{document}